\documentclass[mathleft
]{an}
\usepackage{graphicx}
\usepackage{times}
\begin{document}

% The following seven commands are intended for editorial usage and should be ignored by
% the author(s).
\Pagespan{789}{}% Document's page range. 
% If second parameter is left empty, the last page is computed automatically.
\Yearpublication{2006}%
\Yearsubmission{2005}%
\Month{11}%   
\Volume{999}%  
\Issue{88}% 
% \DOI{This.is/not.aDOI}% 

\title{Differential Rotation in F Stars}

\author{A. Reiners\inst{1,}\thanks{Emmy Noether Fellow}}
%Example 
%for footnote, note the usage of the \texttt{fnmsep}
%command as separator between institute number and footnote mark} 
%\and  G.H. Ostwriter\inst{2,3}

\titlerunning{Differential Rotation in F Stars}
\authorrunning{A. Reiners}
\institute{Georg-August-Universit\"at, Institut f\"ur Astrophysik, Friedrich-Hund-Platz 1, D-37077 G\"ottingen}

\received{\today}
\accepted{\today}
\publonline{later}


\abstract{%
  Differential rotation can be detected in single line profiles of
  stars rotating more rapidly than about $v\,\sin{i} =
  10$\,km\,s$^{-1}$ with the Fourier transform technique. This allows
  to search for differential rotation in large samples to look for
  correlations between differential rotation and other stellar
  parameters. I analyze the fraction of differentially rotating stars
  as a function of color, rotation, and activity in a large sample of
  F-type stars. Color and rotation exhibit a correlation with
  differential rotation in the sense that more stars are rotating
  differentially in the cooler, less rapidly rotating stars. Effects
  of rotation and color, however, cannot be disentangled in the
  underlying sample. No trend with activity is found.  }

\maketitle

\section{Introduction}

Stars are born from molecular clouds carrying net angular momentum
that makes every star rotate more or less rapidly. During their
evolution stars are being braked more or less efficiently, but even
after several Gyrs of efficient braking stars as strongly braked as
the Sun still show substantial angular velocity. This rotation in
general must be expected to be differential, i.e. angular velocity
changes with depth and latitude. In a rotating star, it is already the
temperature difference due to rotationally induced surface gravity
gradients which leads to meridional flow and differential rotation.
More severe effects like magnetic forces and inhomogeneities in the
convective structure may lead to even more severe effects of
differential rotation.

The star with the best studied rotation law is the Sun, its equatorial
angular velocity is roughly 20\,\% higher than the polar one, which
has been observed centuries ago by following the rotation of spot
groups on the solar surface. Today it is believed that these spots are
due to magnetic activity of which at least the cyclic part is
generated by a dynamo mechanism requiring the shear that is caused by
differential rotation.

Detecting differential rotation on stars other than the Sun
unfortunately is not as straightforward as on our host star. Other
stars cannot (yet) be spatially resolved so that we cannot just follow
the latitude-dependent motion of spot groups -- on some stars such
spots might even be absent. Instead, one has to apply indirect
techniques. One way to detect differential rotation is the
reconstruction of the stellar surface using Doppler Imaging (DI) and
comparing the spot's migration at different epochs. This method was
applied successfully to many objects, and a summary of the state of
the art is presented by Collier Cameron in this volume. DI correlates
spot configurations between different epochs that can in principle be
temporally separated as long as the lifetime of the spots, i.e. on the
order of month in the Sun. Hence DI allows the detection of very small
angular velocity differences. This method, however, requires that for
each epoch the star is observed for a full rotation period in order to
reconstruct a good picture of its surface. Depending on the rotation
period and the brightness of the target that requires large amounts of
telescope time, which in case of faint targets have to be quite big to
ensure the small exposure times necessary for a good resolution of the
surface.

A different approach to detect differential rotation is to scrutinize
the shape of the rotationally broadened line profiles, which yields
characteristic differences between rigid and differential rotation.
This method was pioneered in the seventies (Gray, 1977). The problem
of the degeneracy of differential rotation, limb darkening and
inclination was investigated by several authors, (see Reiners \&
Schmitt 2002a and references therein). These authors also provide a
recipe how to detect differential rotation in stellar line profiles. 

Since then, differential rotation was searched for in more than a
thousand spectra of stars from spectral types A--G. This resulted in
many detections of differential rotation. The results of this project
are published in Reiners \& Schmitt (2003a, 2003b), Reiners \& Royer
(2004), and Reiners (2006).  In this paper, I summarize the results
concentrating on the dependency of differential rotation on
temperature, rotation, and activity. I focus on the F-stars, i.e. I
will take into account data from different samples for stars of colors
$0.2 < B-V < 0.6$.

\section{Data sample}

\begin{figure*}
  \centering 
  \mbox{
    \resizebox{.45\hsize}{!}{\includegraphics[angle=-90,bbllx=28,bblly=50,bburx=260,bbury=630]{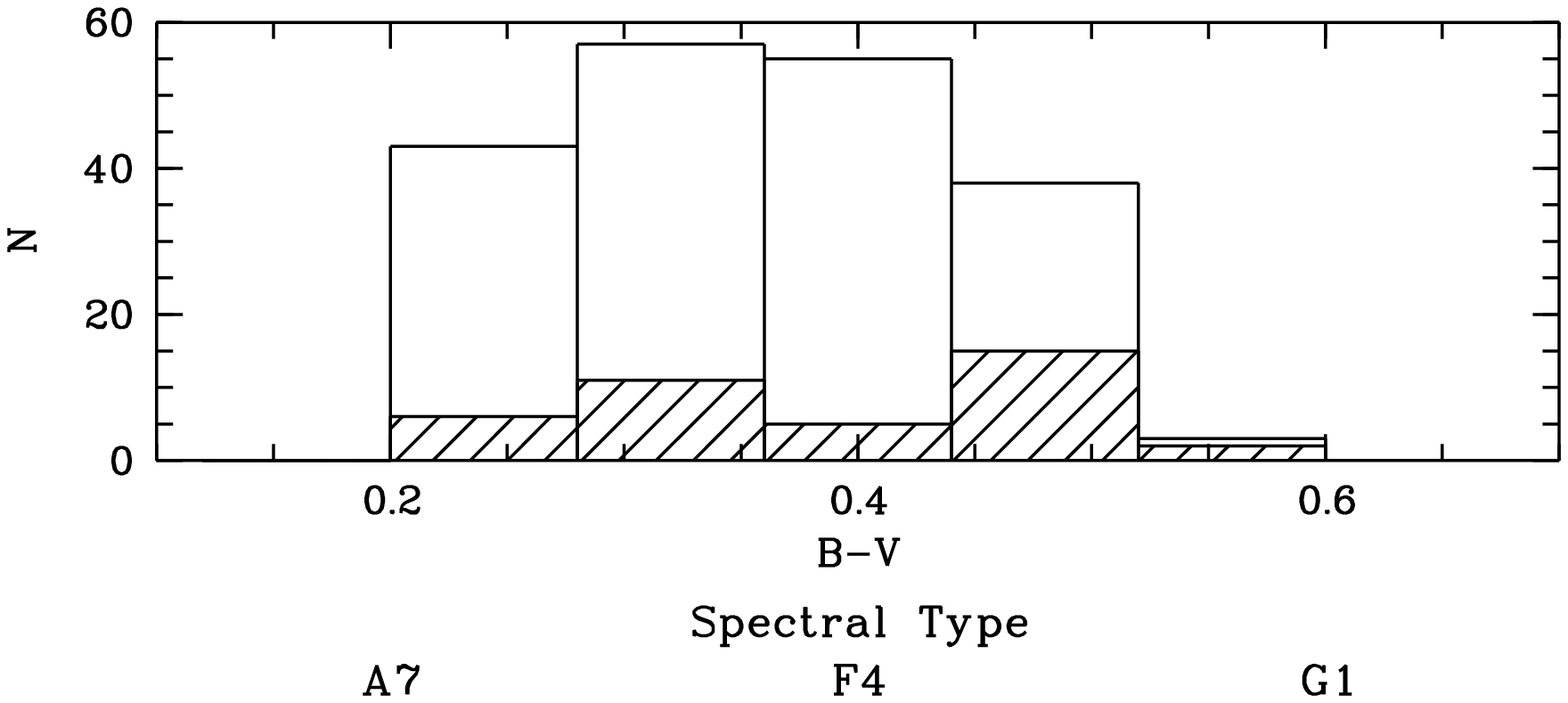}}\hspace{0.05\hsize}
    \resizebox{.45\hsize}{!}{\includegraphics[angle=-90,bbllx=28,bblly=50,bburx=260,bbury=630]{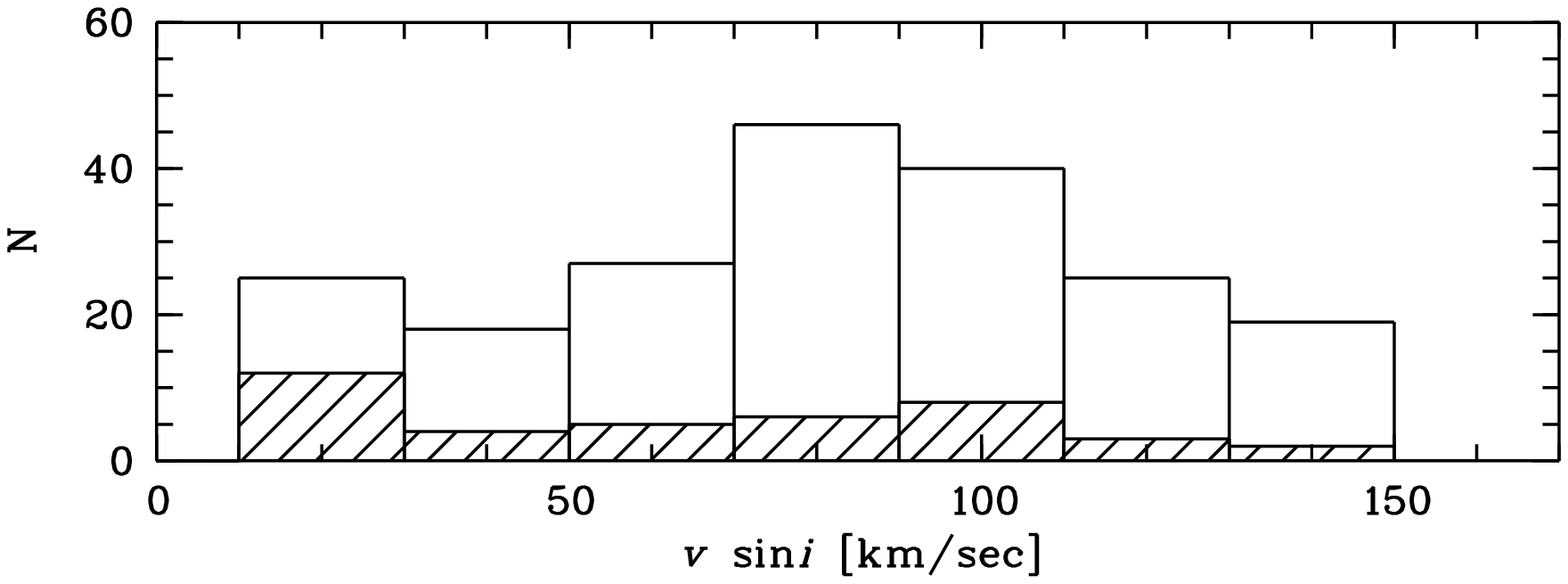}}}
  \mbox{
    \resizebox{.45\hsize}{!}{\includegraphics[bbllx=0,bblly=0,bburx=480,bbury=275]{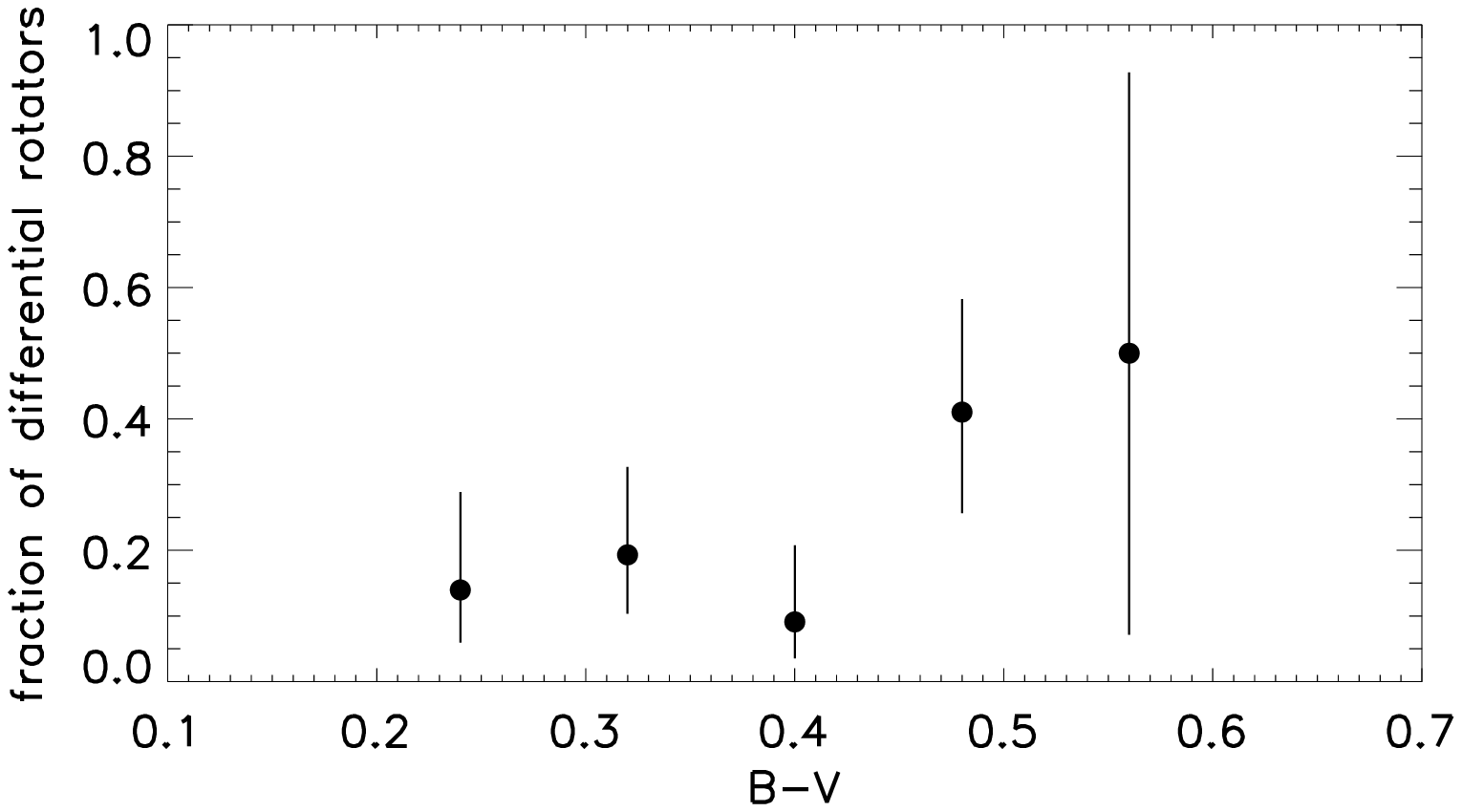}}\hspace{.05\hsize}
    \resizebox{.45\hsize}{!}{\includegraphics[bbllx=0,bblly=0,bburx=615,bbury=340]{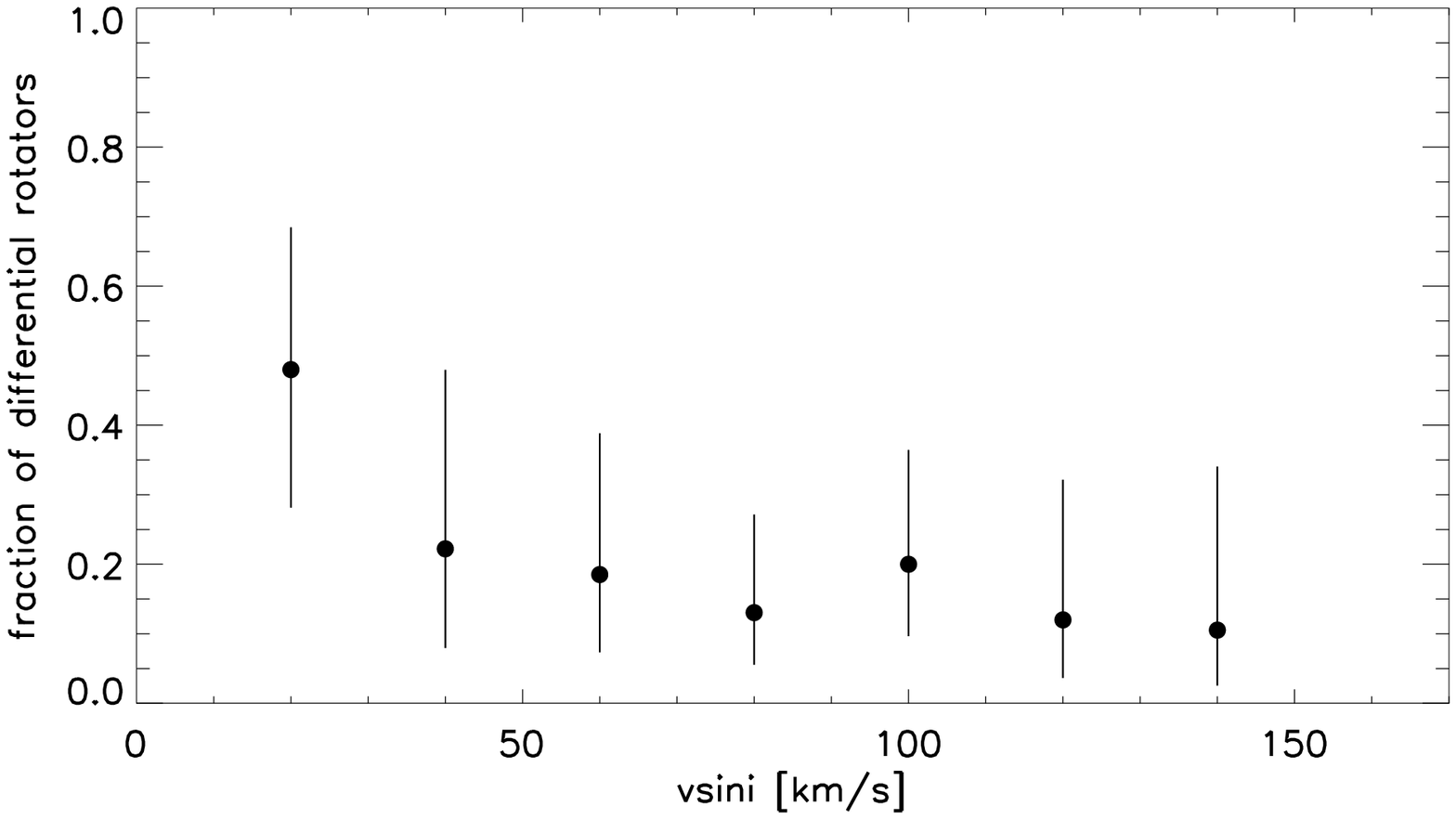}}}
  \caption{\label{plot:histo}Upper panel: Histograms of the sample in $B-V$ (left column)
    and $v\,\sin{i}$ (right column). Lower panel: Fraction of stars
    with strong relative differential rotation in the bins plotted in
    the upper panel.  $2\sigma$-errors are overplotted.}
\end{figure*}

Data are taken from Reiners \& Schmitt (2003a, 2003b), Reiners \&
Royer (2004), and Reiners (2006). The spectra were obtained with FOCES
(CAHA\footnote{Centro Astr\'onomico Hispano Al\'eman}), FEROS, FLAMES,
CES and ECHELEC (ESO). In order to measure the rotation law from a
line profile of a star, it is required that the profile is not
distorted by features due to spots. Hence stars that show asymmetries
in their line profiles are neglected, in their spectra it is not clear
whether line profile distortions are due to the rotation law or
surface features. It should be mentioned that also the symmetric line
profiles could be affected by spots, but the probability of a yielding
a symmetric line profile in the presence of spots is rather low (see
Reiners \& Schmitt, 2002b). An exception to this are polar spots that
always lead to symmetric profiles, these are discussed in
Section\,\ref{sect:polar}.

A total number of 200 stars with colors $0.2 < B-V < 0.6$ yielded
symmetric profiles from which the rotation law could be determined.
Many of the properties of differential rotation were presented in
Reiners (2006) and should not be repeated here. In the following, I
will focus on the stars that do show differential rotation at all, and
which parameters affect or are affected by the presence of strong
latitudinal shear.

\section{What parameters affect differential rotation?}

The measurement of differential rotation is carried out in the
deconvolved broadening function. From that profile, a Fourier
transform is calculated in which the ratio of the second to the first
zero is determined. This ratio is indicative of solar-like
differential rotation (see Reiners \& Schmitt, 2002a). However,
depending on data quality and spectral type, i.e., the number of
useful lines, this ratio has also an uncertainty. In the following, I
will classify a star as differentially rotating if the ratio $q_2/q_1$
indicates differential rotation regardless of how significant this
``detection'' is (see Reiners, 2006). The exact values of $q_2/q_1$
and their uncertainty can be found in the references given above.

\subsection{Effective Temperature}

The upper left panel of Fig.\,\ref{plot:histo} shows a histogram of
the full sample in color $B-V$, the subsample of differential rotators
is also plotted as a hatched histogram. In the lower left panel, the
fraction of stars with signs of differential rotation is shown as a
function of color. As mentioned in Reiners (2006), there is a clear
indication that cooler objects tend to show differential rotation more
frequently. The difference to Fig.\,3 in that paper is that about 50
more early F-stars are added from the sample of Reiners \& Royer
(2004). This leads to much better statistics in the early bins, but it
does not alter the results from Reiners (2006).

\subsection{Rotation}

The right panel of Fig.\,\ref{plot:histo} shows a histogram of the
sample in $v\,\sin{i}$ in the upper panel and the fraction of
differentially rotating stars as a function of $v\,\sin{i}$ in the
lower panel. Again, the conclusions from Reiners (2006) are still
valid; detections of differential rotation are more frequent in slowly
rotating objects.

The observation that differentially rotating stars are more frequently
found among the late stars and among the slow rotators is a
significant result from line profile analysis of several hundred F
stars. However, both results are not independent since in the sample
the early stars tend to be more rapidly rotating, and cooler stars are
in general much slower. Thus, blue $B-V$ color and rapid rotation are
the same in a statistical sense so that the left and right lower
panels in Fig.\,\ref{plot:histo} essentially show the same fact.
Whether it is rotation, temperature, or both that is responsible for
more frequent differential rotation cannot be extracted from this
sample.

\subsection{Activity}

\begin{figure}
  \centering 
    \resizebox{\hsize}{!}{\includegraphics[angle=-90,bbllx=28,bblly=50,bburx=240,bbury=630,clip]{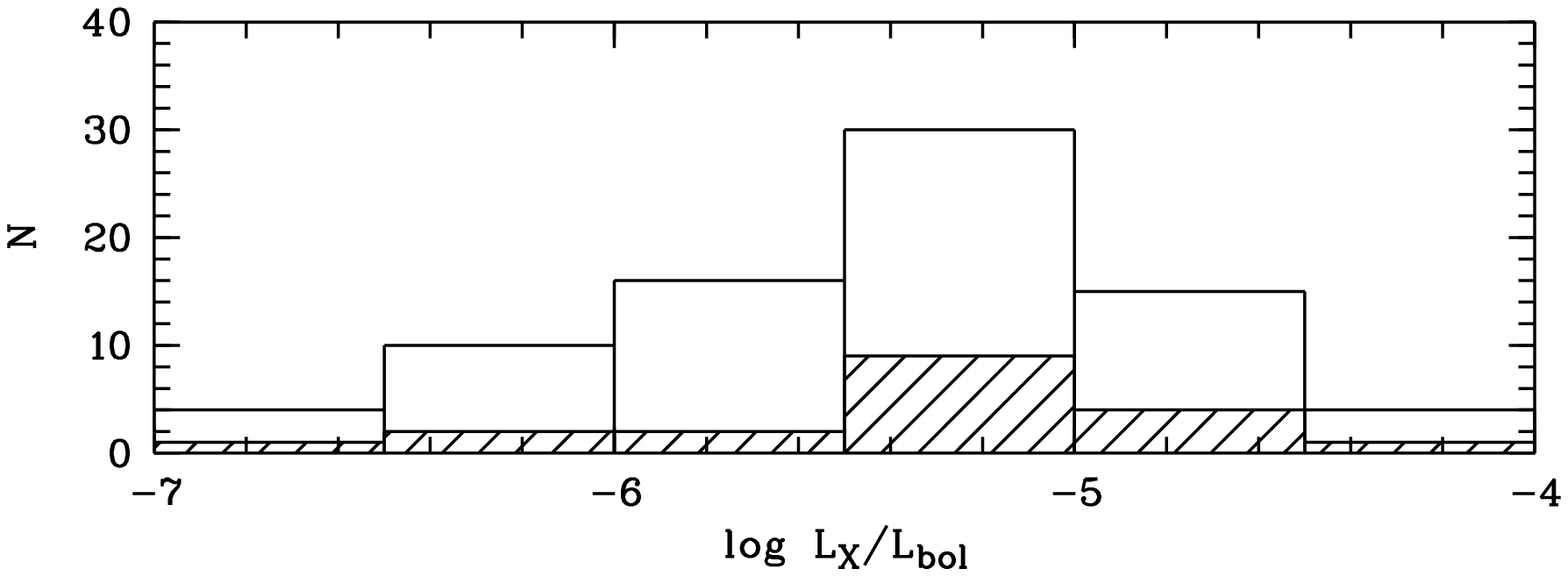}}
    \resizebox{\hsize}{!}{\includegraphics[bbllx=0,bblly=0,bburx=480,bbury=275,clip]{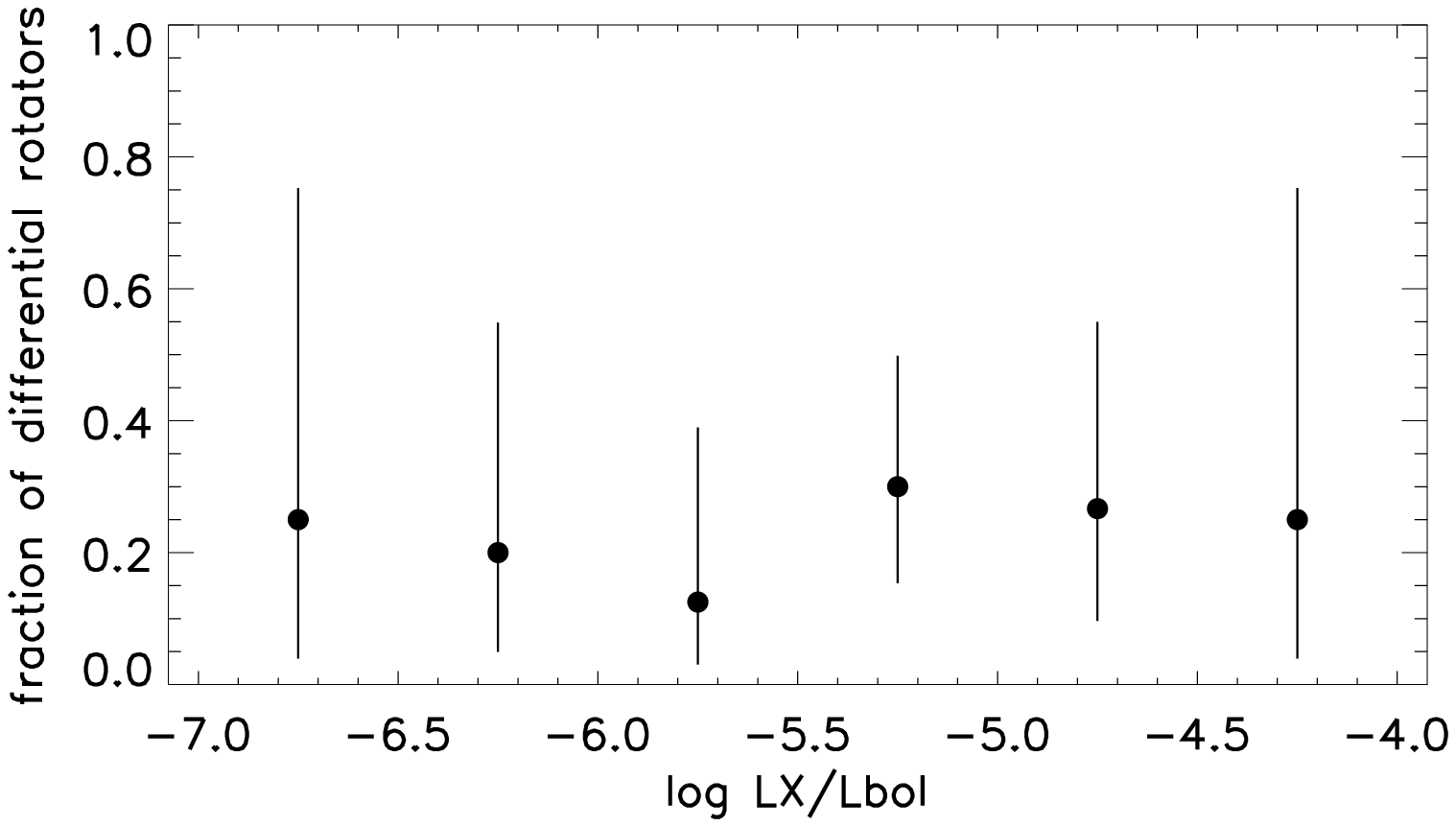}}
    \caption{\label{plot:histo_lxlbol}Upper panel: Histograms of the
      sample in $\log{L_\mathrm{X}/L_\mathrm{bol}}$.  Lower panel:
      Fraction of stars with strong relative differential rotation in
      the bins plotted in the upper panel.  $2\sigma$-errors are
      overplotted.}
\end{figure}

In the standard model of the $\alpha\Omega$ dynamo, radial shear is
required for the generation of magnetic fields and hence for magnetic
activity. If radial shear manifests itself in latitudinal shear at the
stellar surface, this implies that magnetic activity scales with
differential rotation. Observations of X-rays are a good tracer of
magnetic activity. Measurements of X-ray emission are available for
many stars from the ROSAT mission (H\"unsch et al., 1998, 1999).

In the upper panel of Fig.\,\ref{plot:histo_lxlbol}, I show the
distribution of the sample stars for which X-ray measurements are
available as a function of normalized X-ray luminosity
$\log{L_\mathrm{X}/L_\mathrm{bol}}$. In the lower panel, the fraction
of differentially rotating stars is shown as a function of
$\log{L_\mathrm{X}/L_\mathrm{bol}}$. Differential rotation is found in
stars occupying the full range of activity levels contained in the
sample ($-7 < \log{L_\mathrm{X}/L_\mathrm{bol}} < -4$), but the
fraction of differentially rotating stars shows no trend with
activity in the available sample. Although the uncertainties are
still quite large, this already suggests that activity does not
require a strong degree of latitudinal differential rotation, and that
strong differential rotation not necessarily leads to strong activity.

\section{Negative differential rotation}
\label{sect:polar}

Differential rotation might also be anti-solar like, i.e., the equator
rotating at lower angular velocity than the polar regions. Obviously,
this effect does influence stellar line broadening as well and could
be detected in the presented data sample. In fact, there are a number
of spectra that do show the observational signatures of anti-solar
like differential rotation. However, the observational signature is
identical to the signature of cool polar spots (see Reiners \&
Schmitt, 2002b), so that this relatively likely scenario (Schrijver \&
Title, 2001) may be the more realistic explanation. Nevertheless,
negative differential rotation cannot be excluded in a couple of F
stars.

\section{Summary and outlook}

I have presented the results from the analysis of several hundred high
resolution spectra looking for differential rotation in F-stars. The
fraction of stars with signatures of differential rotation in the
overall sample of 200 stars is shown as a function of $B-V$ color,
projected rotation velocity $v\,\sin{i}$, and normalized X-ray
activity $\log{L_\mathrm{X}/L_\mathrm{bol}}$. The fraction of
differential rotators shows a clear trend in color and rotation
velocity, two parameters which unfortunately are not uncorrelated in
the sample. So far the conclusion is that among the cooler, less
rapidly rotating objects the fraction of differentially rotating stars
is larger than among the hotter, more rapidly rotating stars. No trend
is seen in the fraction of differential rotators as a function of
activity.

From the current sample, it cannot be decided whether slow rotation,
low temperature, or both conditions together drive the high fraction
of differentially rotating stars. To decide this, observations in a
well defined sample containing more slowly rotating early F-stars as
well as more rapidly rotating late F-stars are required.

\acknowledgements A.R. acknowledges financial support from the DFG
through an Emmy Noether Fellowship (RE 1664/4-1).

%%%%%%%%%%%%%%%%%%%%%%%%%%%%%%%%%%%%%%%%%%%%%%%%%%%%%%

\end{document}